\documentstyle[preprint,version2,aps]{revtex}
\draft
\begin{document}
\begin{title}
Correlation Induced Insulator to Metal Transitions
\end{title}

\author{Qimiao Si, M. J. Rozenberg, G. Kotliar, and A. E. Ruckenstein}

\begin{instit}
Serin Physics Laboratory, Rutgers University,
Piscataway, NJ 08855-0849, USA
\end{instit}

\centerline{(August, 1993)}

\begin{abstract}

We study a spinless two-band model at half-filling in the limit of
infinite dimensions. The ground state of this model in the
non-interacting limit is a band-insulator. We identify transitions to
a metal and to a charge-Mott insulator, using a combination of
analytical, Quantum Monte Carlo, and zero temperature recursion
methods. The metallic phase is a non-Fermi liquid state with algebraic
local correlation functions with universal exponents over a range of
parameters.

\end{abstract}

\newpage

Strong correlations can lead to low energy behavior qualitatively
different from that predicted by band theory.
A classic example is provided by the Mott transition where a system,
which is expected to be a band metal, turns into an insulator  as
correlations become greater than a critical value.\cite{Mott}
Recently, the complementary question of whether interactions can drive
a band insulator into a metal has been raised in connection with the
possible (semi)-metallic nature of the ``Kondo
insulators''.\cite{Fisk} From a theoretical point of view, it is
likely that a metal which exists only as the result of interactions
can not be analytically continued to a free electron metallic system
and hence should not be described by Fermi liquid theory. The related
question of the possible existence of non-Fermi liquid metallic states
has been of great recent interest, due to the anomalous normal state
properties of the high-$T_c$ oxides~\cite{Anderson,MFL}.

In this letter, we study a spinless model at half-filling which
displays a band insulating ground state  in the absence of
interactions. We developed two numerical algorithms to study the
the effects of interactions in infinite dimensions, supplementing the
numerical results with analytical arguments. We find that, in spite of
the simplicity of the model, the interactions give rise to a novel
correlation-induced insulator (to which we refer as a charge-Mott
insulator), and a non-Fermi liquid metallic phase with scale invariant
low-energy correlation functions. In the latter, we identify two
parameter regimes, in which the associated exponents are continuously
varying and universal, respectively.

We study the following Hamiltonian

\begin{eqnarray}
{\rm H} = &&\sum_{k} (\epsilon_{\bf k} -\mu) c_{\bf k}^{\dagger} c_{\bf k}
+ \sum_i  (
{\epsilon}^o_d -\mu) d^{\dagger}_{i} d_{i}\nonumber\\
&&+\sum_i  t ( d^{\dagger}_{i}
c_{i} + h.c. ) + {V} \sum_i
 (c^{\dagger}_{i} c_{i} - {1 \over 2}) (d^{\dagger}_{i } d_{i} -{1 \over 2}) ,
\label{hamiltonian}
\end{eqnarray}
where  $c^{\dagger}$ and $d^{\dagger}$ are creation operators for two
species of spinless electrons.  The $c$-electrons have a dispersion
$\epsilon_k$,
while the $d$-band is dispersionless with local  energy
$\epsilon_d^o$. These two species are coupled through the
hybridization, $t$, and the density-density interaction, $V$. Below we
focus on the half-filled case, with $\epsilon_d^o=\epsilon_c=\mu=0$.

We solve this model in the limit of infinite
dimensions\cite{Vollhardt} in which
the hopping matrix element $t_{ij}$ must be rescaled with
$1/\sqrt{d}$. The one-particle Green's function, $G ({\bf k},
i\omega_n) \equiv -\int d \tau {\rm e}^{i\omega_n \tau} <T_{\tau}
\psi_{\alpha} ({\bf k},\tau ) \psi_{\alpha}^{\dagger}( {\bf k},\tau
)>$, with  $\psi^\dagger = ( d^\dagger , c^\dagger )$, can be
calculated from the Dyson's equation

\begin{eqnarray}
G^{-1}({\bf k}, i\omega_n) ~\equiv~
\left( \begin{array}{ll}
        i\omega_n+\mu-\epsilon_d^o &
{}~ ~ ~ ~ t\\
        ~ ~ ~ ~ t & i\omega_n+\mu-\epsilon_{\bf k} \end{array}
\right) - \Sigma (i\omega_n)
\label{gfdef}
\end{eqnarray}
The self-energy $\Sigma (i\omega_n )$ is momentum independent and can
thus be calculated from the local Green function, $<\psi ^{\dagger}
\psi>_{\rm S_{\rm imp}}$, associated with the impurity
action~\cite{GK},

\begin{eqnarray}
{\rm S}_{\rm imp} (G_o)= &&-\int^{\beta}_0 d\tau d\tau '
\psi^{\dagger}(\tau )G^{-1}_o (\tau - \tau' ) \psi ( \tau'
) + \int^{\beta}_0 d\tau {V} (c^{\dagger} c -{1 \over 2}) (
d^{\dagger} d - {1 \over 2})
\label{siteaction}
\end{eqnarray}
according to the Dyson equation, $\Sigma =
G_o^{-1}-<\psi^{\dagger} \psi >_{\rm S_{\rm imp}}$, where
$G_o$ obeys the self-consistency equation
$\sum_{\bf k} G ({\bf k}, i\omega_n) /N_{site} =
<\psi^{\dagger} \psi >_{\rm S_{\rm imp}}$. As is already implicit in
the choice of the local action (\ref{siteaction}), we will focus on
states with no long range (CDW) order\cite{note:cdw}.

The effective-impurity action (\ref{siteaction}) can be equivalently
written in terms of a resonant-level model,

\begin{eqnarray}
{\rm H}_{\rm imp} = {\rm H}_o + t(c^{\dagger}d + d^{\dagger}c ) +
(\epsilon_d^o-\mu) d^{\dagger}d + V (n_d -{1 \over 2}) (n_c -{1 \over 2})
\label{rlm}
\end{eqnarray}
where ${\rm H_o}$ describes the local $c$-electron coupled to a
non-interacting electron bath with a hopping matrix determined from
the self-consistency equation. For simplicity we study the case of a
semi-circular conduction-electron density of states, $\rho_o
(\epsilon) = \sum_{\bf k}\delta(\epsilon-\epsilon_{\bf k})/ N_{site}
= (2 / {\pi D}) \sqrt{1 - ( \epsilon /D)^2}$, for which the
self-consistency equation reduces to the form:

\begin{eqnarray}
(G_o^{-1})_{cc} = i\omega_n + \mu -({D \over 2})^2G_{cc}
\label{consistency}
\end{eqnarray}

We developed two numerical techniques to
analyze Eqs. (\ref{siteaction}-\ref{consistency}). The first
algorithm is a generalization of the method used in previous studies
of the Hubbard model.\cite{Jarrell,Rozenberg,Georges} An initial guess
for $G_o$ is used in the impurity action Eq. (\ref{siteaction}), which
is solved by Monte Carlo method. Our algorithm is based on a
modification of that due to Hirsch and Fye\cite{Hirsch}, and uses
a different set of Hubbard-Stratonovich variables~\cite{hs} and a {\it
single} determinant. The calculated Green's function is then fed into
the self-consistency condition (\ref{consistency}) and the procedure
is iterated until convergence is achieved.

The second algorithm is based on the recursion
method.\cite{Haydoch,Lanczos} The one particle Green's function can be
written as $G(z)={G^>}(z)+{G^<}(z)$ where $G^{> \atop <} (z) = \int d
\epsilon {\rho_{\pm} (\epsilon ) / (z-\epsilon})$ and $\rho_{\pm}
(\epsilon ) =0$ for $\epsilon {> \atop <} 0$. We use the
continued-fraction expression for $G^{> }(z)$,

\begin{eqnarray}
G^{>} (z) = {b_1^r \over \displaystyle z - a_1^r - {\strut b_2^r \over
\displaystyle z -a_2^r-{\strut b_3^r \over {\cdots}}}},
\label{cf}
\end{eqnarray}
and a similar one for $G^{<} (z)$, with $a_i^r$ and $b_i^r$ replaced
by $a_i^l$ and $b_i^l$. This representation obeys the correct
analyticity requirements ($b_1^r+b_1^l=1$). It allows us to visualize
the self-consistency equation, $(G_o^{-1})_{cc} = i \omega_n -
(\epsilon_c-\mu) -(D/2)^2G_{cc}$, as the inverse of the local Green's
function $<c^{\dagger} c > $ of the non-interacting Hamiltonian,

\begin{eqnarray}
{\rm H}_o=(\epsilon_c - \mu ) c^{\dagger}c
&&+{D\over 2}\sqrt{b_1^r}(c^{\dagger}c_{1r} +h.c. )+ \sum_{i=1}^N a_i^r
c_{ir}^{\dagger}c_{ir} +\sum_{i=1}^{N-1} \sqrt{b_{i+1}^r} (
c_{ir}^{\dagger}c_{i+1 r} +h.c. )\nonumber\\
&&+{D\over 2}\sqrt{b_1^l}(c^{\dagger}c_{1 l} +h.c. )+ \sum_{i=1}^N a_i^l
c_{il}^{\dagger}c_{il} +\sum_{i=1}^{N-1} \sqrt{b_{i+1}^l} (
c_{il}^{\dagger}c_{i+1 l} +h.c. )
\label{chain}
\end{eqnarray}
where $N$ is the level of the continued-fraction expression which
is infinite in the exact representation but taken as finite in
practice. The problem can now be reformulated in terms of the
self-consistent determination of the
variables $a_i^r$, $a_i^l$, $b_i^r$, and $b_i^l$\cite{Werner}.
We start with an initial guess for these parameters, and solve the
ground state $|0>$ of the
resonant-level Hamiltonian given by (\ref{rlm}) and (\ref{chain})
using the standard sparse matrix diagonalization
methods.\cite{Lanczos} The new parameters $a_i^r$, $a_i^l$, $b_i^r$,
and $b_i^l$ for the next iteration is then obtained from the matrix
elements of an orthogonal basis which tridiagonalizes ${\rm H}_{\rm
imp}$. Specifically, $a_{n}^r = <f_{n}^r|{\rm H}_{\rm
imp}|f_{n}^r>/<f_{n}^r|f_{n}^r>$, and
$b_{n}^r=<f_n^r|f_n^r>/<f_{n-1}^r|f_{n-1}^r>$, $b_1^r=<f_1^r|f_1^r>$,
where $|f_1^r>=c|0>$, $|f_{2}^r>={\rm H}_{\rm
imp}|f_1^r>-a_1^r|f_1^r>$ and for $n\ge 2$, $|f_{n+1}^r>={\rm H}_{\rm
imp}|f_n^r>-a_{n}^r|f_n^r>-b_n^r|f_{n-1}^r>$. Similar equations, with
$|f_1^l>=c^\dagger |0>$, give rise to $a_i^l$ and $b_i^l$. Compared to
the finite temperature Monte Carlo method, this algorithm works
directly at zero temperature, and is less expansive numerically.

Using both methods, we find that the model displays two
kinds of insulating phases separated by a metallic phase. The  phase
diagram in the $V-t$ parameter space is shown schematically in Fig.
\ref{pd}. To illustrate the phase diagram, we plot in Fig. \ref{ins}
the $d$-electron local Green's function, $G_{dd}$, as a function of
the Matsubara frequency for two parameters with a same value of the
hybridization, $t=0.4D$,
and different values of the interaction, $V=-0.1D$ and $V=-2.5D$,
respectively. They clearly display the gap in their respective single
particle excitation spectrum, and also demonstrate the excellent
agreement between the zero temperature results from the recursion
method and the low (but finite) temperature results from the Quantum
Monte Carlo. Shown in Fig. \ref{metal} are $G_{dd}$, as well as the
$c$-electron local Green's function, $G_{cc}$, for the hybridization
identical to that used in Fig. \ref{ins}, $t=0.4D$, and an interaction
strength in between those used in Fig. \ref{ins}, $V=-0.9D$. The
single-particle excitation spectrum is gapless in this case.

In order to clarify the nature of the phases and the structure of the
phase diagram, we now consider various limiting cases where analytical
arguments can be made. In the non-interacting limit ($V=0$)  $\Sigma
=0$ and  the Green function (\ref{gfdef}) displays two branches of
poles, $E_{\bf k}^{\pm} = (\epsilon_{\bf k} \pm \sqrt { (\epsilon_{\bf
k})^2 + 4 t^2 })/2$, describing the fully occupied bonding band ($-$)
and the empty anti-bonding band ($+$). Thus, a hybridization gap
$\Delta_o = \sqrt {D^2+4 t^2}-D$, develops, leading to band-insulating
behavior. This state becomes unstable whenever the effect of
interactions is strong enough to close the hybridization gap.

This can be most easily seen in the atomic limit, $D=0$, in which
case the behavior of the  effective impurity model
(\ref{siteaction}-\ref{consistency}) is determined by the local empty,
doubly occupied, bonding and anti-bonding singly occupied electron
states, respectively,

\begin{eqnarray}
&& |E>=|N=0>=|0>\nonumber\\
&& |D>=|N=2> = d^{\dagger} c^{\dagger}|0> \nonumber\\
&& |B>=|N=1, b> = {1 \over {\sqrt{2}}}(d^{\dagger} -
c^{\dagger})|0>\nonumber\\
&& |A>=|N=1, ab> = {1 \over {\sqrt{2}}}(d^{\dagger} + c^{\dagger})|0>,
\label{atomic-states}
\end{eqnarray}
The corresponding eigenvalues are $\epsilon (E) = \epsilon (D)= V/4$,
$\epsilon (B) = -V/4 -t$, and $\epsilon (A) = -V/4 +t$. Three
parameter regimes occur in this limit:  (i) For $-V < 2t$ the local
bonding state,  $|B>$, has the lowest energy, and is separated by a
finite gap from lowest single particle excitation, leading to
band-insulating behavior; (ii) At $-V=2t$ three states, $|B>$, $|D>$,
and $|E>$, become degenerate and metallic behavior with gapless
single-particle excitations sets in; (iii) Finally, for $-V > 2t$ the
low-energy charge doublet,  $|D>$ and $|E>$, is separated by a gap
from the lowest single-particle excitation leading to a situation
analogous to the Mott insulator of the atomic limit of the half-filled
repulsive Hubbard model, in which case the spin doublet is the lowest
energy state.\cite{Rozenberg} We thus refer to the resulting phase as
a charge-Mott insulator.\cite{cmi}

The stability of these three phases and the position of the boundaries
between them, can also be understood analytically in the limit of
small hybridization.  In that case, the instability of the band
insulator, which occurs through the closing of the hybridization gap,
is most easily seen in terms of  the {\em irrelevance} of the
hybridization coupling (in the sense of the renormalization group) for
a sufficiently large value of the attractive interaction, $V$.
The value of the critical coupling, $V_{c1}$, follows from a scaling
analysis of the resonant-level model (\ref{rlm}), in the vicinity of
the phase boundary from the metallic ($t\approx 0$) side. It is
obtained from the condition, $\epsilon _t =1$, on the the anomalous
dimension of the hybridization operator, $\epsilon _t$, defined from
the long time behavior of the auto-correlation function, $\chi (\tau)
=- <T_{\tau} d^{\dagger} ( \tau ) c ( \tau ) c^{\dagger}(0)d(0)> \sim
\tau^{-2\epsilon_t}$. Such a condition implies~\cite{SK} $\rho
(V_{c1}) V_{c1} = -(2/{\pi}) tan[\pi (\sqrt{2} -1)/2]$. Here, $\rho
(V) = (2 / \pi D)/ \sqrt{1-({V / D})^2}$ is the bath density of states
at the Fermi level which is itself dependent on $V$ as a result of the
self-consistency requirement (\ref{consistency}).\cite{note:bathdos}
This argument leads to $V_{c1} = - 0.6D$. Upon further increasing the
attraction beyond $-V_{c1}$ the charge-Mott gap develops at a second
critical value, $-V_{c2}$. The latter can be  calculated from the
exact solution of the Falicov-Kimball model (at
$d=\infty$)~\cite{Brandt} corresponding to the Hamiltonian
(\ref{hamiltonian}) with $t=0$, and is identified as the value of $V$
at which the local conduction electron density of states at the Fermi
energy, $\rho_c(0)$, starts to vanish. For the semi-circular density
of states, $\rho_c (0)=(2 /\pi D)\sqrt{1-({V\over
D})^2},$~\cite{vanDongen} leading to $V_{c2}=-D$. Our numerical
results are consistent with the analytical estimates for $V_{c1}$ and
$V_{c2}$.

To further characterize the nature of the metallic phase, we have
analyzed numerically the low frequency behavior of the local Green's
functions. Consider first the limit of small hybridizations. In this
regime, we find that $G_{d} \sim  -i |\omega_n|^{-\alpha} sgn
\omega_n$.
This self-similar behavior is consistent with the irrelevant
hybridization , which implies a zero renormalized Fermi
energy.\cite{SK}  In this case, the exponent is expected to to be
given by the orthogonality exponent, $({\delta \over\pi})^2$, with the
phase shift determined from the standard formula, $\delta = (2/\pi)
tan ^{-1} (\pi \rho (V) V /2)$.
The numerically determined exponents  are consistent with this
analytical expression.
Finally, the calculated $c$-electron local Green's function, $G_{cc}$,
is nonsingular, as expected from a picture of the metallic state in
which the local degrees of freedom asymptotically decouple from the
conduction-electron bath at low energies.\cite{SK,SKG}

The most interesting regime of the metallic phase occurs for
intermediate to large values of the hybridization (as well as the
interaction strength). In this regime, our numerical results show
that, $G_{dd}$ continues to have a self-similar behavior,
with an exponent $\alpha$ close to 1, independent of the interaction
strength.
At the same time, $G_{cc}$ is again nonsingular, suggesting that, even
in the intermediate to large hybridization limit (for finite
bandwidth, $2D$), a partial decoupling of the $d$ and $c$ electron
degrees of freedom occurs at low energies.

Such a partial decoupling of the local degrees of freedom, and the
interaction-independence of the exponents, can be  understood as a
result of the self-consistent modification of the electron bath for
the effective impurity model using the atomic limit metallic state as
a starting point. In that case, projecting onto the three degenerate
low-energy atomic states, $|B>$, $|D>$, and $|E>$, in
(\ref{atomic-states}), leads to the representation of local fermions,
$c^{\dagger}=-{1\over \sqrt{2}}(X_{BE}+X_{DB})$ and
$d^{\dagger}={1\over \sqrt{2}}(X_{BE}-X_{DB})$ in terms of  projection
operators, $X_{\alpha \beta} \equiv |\alpha><\beta|$.  The
corresponding low-energy spectral functions take the form,
$\rho_c(\omega) = \rho_d (\omega )={2 \over 3} \delta (\omega )$. In
the presence of the hopping term $D$, these atomic configurations are
coupled to the self-consistent electron bath through the $H_o$ term in
(\ref{rlm}) which we parametrize as,
$H_o=\sum_kV_k(c^{\dagger}\eta_k+H.c.)+
\sum_k\tilde{\epsilon_k}\eta_k^{\dagger}\eta_k$. The parameters,
$\tilde \epsilon _k$ and $V_k$, are determined from the
self-consistency condition,
$(1/N_{site})\sum_kV_k^2\delta(\epsilon-\tilde{\epsilon}_k) = (D/2)^2
\rho_c (\epsilon)$, which corresponds to taking the imaginary part of
Eq. (\ref{consistency}). The form of $\rho_c$ in the atomic limit
implies that the electron bath $\eta_k$ corresponds to a single level,
created by $\eta^{\dagger} _o$, at zero energy. Hence, the impurity
problem reduces to $\tilde {\rm H}_{\rm imp} = -{D \over 2\sqrt{3}}
[(X_{BE}+X_{DB})\eta _o +h.c.]$, which acts on a Hilbert space
spanned by $|E>$, $|B>$,$|D>$ and $\eta^{\dagger} _o |E>$,
$\eta^{\dagger} _o|B>$,$\eta^{\dagger} _o |D>$. It is easily seen that
the ground state
manifold of this Hamiltonian is doubly degenerate and consists of the
states $|N=1>=(1 /{\sqrt{2}}) (|B>+\eta^{\dagger} _o |E>)$, and
$|N=2>=(1/{\sqrt{2}}) (|D>+\eta^{\dagger} _o |B>)$.  Recalculating the
Green functions of local fermions we obtain a zero energy pole in
$G_{dd}$ and pairs of non-zero energy poles, symmetrically centered
around zero, in $G_{cc}$. This form of $G_{cc}$ then implies that the
electron bath for the next iteration is composed of pairs of levels at
energies $\pm \epsilon_i$. It can be shown, following the same
procedure, that for each pair of levels at energies ${\pm \epsilon}$
coupled to the local $c$-electron, there again exist degenerate ground
states, leading to a zero energy pole in $G_{dd}$ and pairs of
non-zero energy poles in $G_{cc}$. As the iteration proceeds, the
poles in $G_{cc}$ spread over a range of the order of the bare
bandwidth while a zero-energy pole persists in $G_{dd}$, leading to
the self-consistent solution with a regular $G_{cc}$ and a
self-similar $G_{dd}$ with the exponent $\alpha=1$.

In summary, we reported a) novel transitions from a band-insulator to
a metal and to a charge-Mott insulator within a simple spinless
two-band model; and b) a novel non-Fermi liquid metallic regime
characterized by an algebraic local correlation functions with
universal exponents. These local correlation functions are shown
to be associated with a non-trivial effective impurity problem. And
the very existence of this new non-Fermi liquid is closely related to
a level-crossing in the atomic limit of the associated impurity
problem, a benchmark for a class of non-Fermi liquid states in a
variety of strongly correlated impurity
models\cite{multic,twoimp,SK,Perakis}. Related ideas have been
proposed in a phenomenological discussion of strongly correlated
lattice models in which non-Fermi liquid behavior arises due to the
coexistence of single particle states and dispersionless three-body
resonances at the chemical potential.\cite{Andrei} The
results of this letter are expected to have important implications for
models describing realistic strongly correlated systems such as the
``Kondo-Insulators'' and high $T_c$ copper oxides. In this regard we
note that, while the metallic phase in our model occurs in a range of
attractive interactions, these interactions should be understood as
effective parameters derived from
the bare interactions in a more complex set of local orbitals in real
materials.\cite{KS} We also note that, in the spinful cases and away
from half-filling, there can exist an intermediate non-Fermi liquid
metallic state with coherent spin excitations and incoherent charge
excitations.\cite{SK} Our results here should describe the charge
sector of the intermediate phase. These and other related questions
are currently under investigation within more realistic models.

We thank V. Dobrosavljevic, A. Georges, C. Varma, and X.Y. Zhang for
useful discussions. Q.S. and A.E.R. acknowledge the hospitality of the
Aspen Center for Physics where part of the work was completed. This
work was supported by the NSF under grant \#DMR 922-4000 (G.K.) and by
the ONR under grant \# N00014-92-J-1378 (A.E.R.).

\figure{The qualitative phase diagram. Here ``BI'', ``M'', and ``CMI''
correspond to the band insulating, metallic, and charge-Mott
insulating phases respectively. \label{pd}}

\figure{$G_{dd}$ vs. Matsubara frequency $\omega_n$ for $V=-0.1D$ and
$t=0.4D$ (band insulator) calculated from the recursion method (solid
line) using 10 sites to truncate the impurity Hamiltonian and 100
levels for the continued-fraction expansion, and from quantum Monte
Carlo (circles) with $\beta=32$. Similar results are shown (dashed
line and squares) for $V=-2.5D$ and $t=0.4D$ (charge-Mott
insulator).\label{ins}}

\figure{$G_{dd}$ vs. Matsubara frequency $\omega_n$ for $V=-0.9D$
and $t=0.4D$ (metal) calculated from the recursion method (solid line)
with 10 sites and 100 levels and quantum Monte Carlo (circles) with
$\beta=64$. Inset plots the similar results for $G_{cc}$.
\label{metal}}


\begin{references}
\bibitem[16]{cmi} The charge-Mott insulator exhibits Curie-like local
charge and $d^{\dagger}c^{\dagger}$ pairing susceptibilities,
signaling a CDW instability at a finite temperature
$T_c$.\cite{note:cdw} $T_c$ vanishes as $|V| \rightarrow \infty$ and
can be strongly suppressed when a system is frustrated.
\bibitem[18]{note:bathdos} Specifically, $\rho(V)$ is defined as
$-{\rm Im}(G_o)_{cc}(i0^{+})/\pi$. The self-consistency equation
(\ref{consistency}) then yields $\rho(V)=(2/\pi D)^2/\rho_c(0)$.
\end{references}
\end{document}